# The GOES Microburst Windspeed Potential Index

Kenneth L. Pryor
Center for Satellite Applications and Research (NOAA/NESDIS)
Camp Springs, MD

## Abstract

A suite of products has been developed and evaluated to assess hazards presented by convective downbursts to aircraft in flight derived from the current generation of Geostationary Operational Environmental Satellite (GOES) (11-12). The existing suite of GOES microburst products employs the GOES sounder to calculate risk based on conceptual models of favorable environmental profiles for convective downburst generation. A GOES sounder-derived wet microburst severity index (WMSI) product to assess the potential magnitude of convective downbursts, incorporating convective available potential energy (CAPE) as well as the vertical theta-e difference (TeD) between the surface and mid-troposphere has been developed and implemented. CAPE has an important role in precipitation formation due to the strong dependence of updraft strength and resultant storm precipitation content on positive buoyant energy. Intended to supplement the use of the GOES WMSI product over the United States Great Plains region, a GOES Hybrid Microburst Index (HMI) product has also evolved. The HMI product infers the presence of a convective boundary layer by incorporating the sub-cloud temperature lapse rate as well as the dew point depression difference between the typical level of a convective cloud base and the sub-cloud layer. Thus, the WMSI algorithm is designed to parameterize the physical processes of updraft and downdraft generation within the convective storm cell, while the HMI algorithm describes the moisture stratification of the sub-cloud layer that may result in further downdraft acceleration, eventually producing a downburst when the convective downdraft impinges on the earth's surface. It is proposed to merge the WMSI and HMI algorithms into a Microburst Windspeed Potential Index (MWPI) algorithm for implementation in the GOES-R Advanced Baseline Imager (ABI).

## 1. Introduction

A favorable environment for downbursts associated with deep convective storms that occur over the central and eastern continental United States includes strong static instability with large amounts of convective available potential energy (CAPE). Previous research (Fujita 1985, Ellrod 1989) has identified that over the central United States, especially in the Great Plains region, an environment between that favorable for wet microbursts (Atkins and Wakimoto 1991) and dry microbursts (Wakimoto 1985) may exist during the convective season in which sub-cloud evaporation of precipitation is a significant factor in downdraft acceleration. This intermediate type environment, as described by Caracena et al. (2007), is characterized by conditions favorable for both wet and dry microbursts:
1.Significant CAPE.
2.A deep, dry adiabatic lapse rate layer below the cloud base, which is typically near the 700 millibar (mb) level.

A suite of products has been developed and evaluated to assess hazards presented by convective downbursts to aircraft in flight derived from the current generation of Geostationary Operational Environmental Satellite (GOES) (Menzel et al. 1998). The existing suite of GOES microburst products employs the GOES sounder to calculate risk based on conceptual models of favorable environmental profiles for convective downburst generation. Pryor and Ellrod (2004) and Pryor and Ellrod (2005) outlined the development a Geostationary Operational Environmental Satellite (GOES) sounder-derived wet microburst severity index (WMSI) product to assess the potential magnitude of convective downbursts, incorporating CAPE as well as the vertical theta-e difference (TeD) between the surface and mid-troposphere. In addition, Pryor (2006a) developed a GOES Hybrid Microburst Index (HMI) product intended to supplement the use of the GOES WMSI product over the United States Great Plains region. The HMI product infers the presence of a convective boundary layer (CBL) (Stull 1988, Sorbjan 2003) by incorporating the sub-cloud temperature lapse rate as well as the dew point depression difference between the typical level of a warm season Great Plains convective cloud base and the sub-cloud layer. Thus, the WMSI algorithm is designed to parameterize the physical processes of updraft and downdraft generation within the convective storm cell, while the HMI algorithm describes the moisture stratification of the sub-cloud layer that may result in further downdraft acceleration, eventually producing a downburst when the convective downdraft impinges on the earth's surface.  Based on validation of the GOES WMSI and HMI products over the Oklahoma Panhandle during the 2005 and 2006 convective seasons, Pryor (2006b) noted an inverse proportionality between WMSI and HMI values for convective wind gusts of comparable magnitude.  The statistically significant negative correlation between WMSI and HMI values likely reflects a continuum of favorable environments for downbursts, ranging from wet, represented by high WMSI and low HMI values, to an intermediate or hybrid regime, characterized by lower WMSI values and elevated HMI values. The inverse relationship between WMSI and HMI values associated with observed downbursts underscores the relative importance of thermal and moisture stratification of the boundary layer in the acceleration of convective downdrafts and resulting downburst wind gust magnitude.

    Accordingly, the GOES Microburst Windspeed Potential Index (MWPI) algorithm, derived from merging the WMSI and HMI, is designed to infer the presence of a CBL by incorporating the sub-cloud lapse rate between the 670 and 850 mb levels as well as the dew point depression difference between the typical level of a convective cloud base at 670 mb and the sub-cloud layer at 850 mb. In a typical dry microburst thermodynamic environment, Wakimoto (1985) identified a convective cloud base height near the 500 mb level. In contrast, Atkins and Wakimoto (1991) identified a typical cloud base height in a pure wet microburst environment near 850 mb. Thus, an intermediate cloud base height of 670 mb was selected for a hypothetical hybrid microburst environment. This selection agrees well with the mean level of free convection (LFC) of 670 mb computed from the inspection of twenty GOES proximity soundings corresponding to downburst events that occurred in Oklahoma between 1 June and 31 July 2005. In a free convective thermodynamic environment (i.e. no convective inhibition (CIN)), the mean LFC of 670 mb can be considered representative of convective cloud base heights that occur in an environment favorable for hybrid microbursts such that
LFC ≈ LCL ≈ $z_i$ ≈ 670 mb, where LCL is the lifting condensation level and $z_i$ represents mixed layer depth. CAPE has an important role in precipitation formation due to the strong dependence of updraft strength and resultant storm precipitation content on positive buoyant energy. The formation of precipitation and subsequent loading initiate convective downdrafts. CAPE is an important parameter of consideration due to the fact that the precipitation caused by updrafts will produce the sub-cloud evaporational cooling and negative buoyancy that accelerates convective downdrafts. Thus, the Microburst Windspeed Potential Index (MWPI), is defined as

$$\text{MWPI} = (\text{CAPE}/100) + G + (T - T_d)_{850} - (T - T_d)_{670} \qquad (1)$$

where G is the lapse rate in degrees Celsius (C) per kilometer from the 850 to the 670 mb level, T is temperature in degrees Celsius, and $T_d$ is the dewpoint temperature (C).  Climatology of severe storm environmental parameters (Nair et al. 2002) has found that a deeper convective mixed layer, as represented by large LFCs, predominates in the warm season over the southern Plains. The presence of a deep, dry sub-cloud (mixed) layer will enhance evaporational cooling and downdraft intensification as precipitation falls below the convective storm cloud base. In fact, it was found by Nair et al. (2002) that moderately high LFCs, that coexist with large CAPE over the Great Plains, are associated with an observed maximum in severe convective storm occurrence.  Thus, this paper proposes to merge the

WMSI and HMI algorithms into the Microburst Windspeed Potential Index (MWPI) algorithm for implementation in the GOES-R Advanced Baseline Imager (ABI).

## 2. Methodology

Data from the GOES HMI and MWPI was collected over the Oklahoma Panhandle for downburst events that occurred on 29 March and 10 April 2007 and validated against conventional surface data. Images were generated by Man computer Interactive Data Access System (McIDAS) and then archived on an FTP server (ftp://ftp.orbit.nesdis.noaa.gov/pub/smcd/opdb/wmsihmiok/). Cloud-to-ground (CG) lightning data from the National Lightning Detection Network (NLDN) was plotted over GOES imagery to compare spatial patterns of CG lightning to surface observations of downburst wind gusts. The Oklahoma Panhandle was chosen as a study region due to the wealth of surface observation data provided by the Oklahoma Mesonet (Brock et al. 1995), a thermodynamic environment typical of the High Plains region during the warm season, proximity to the dryline, and relatively homogeneous topography. The High Plains region encompasses the Oklahoma Panhandle that extends from 100 to 103 degrees West (W) longitude and is characterized by short-grass prairie. The treeless, low-relief topography that dominates the sparsely populated Oklahoma Panhandle allows for the assumption of horizontal homogeneity when deriving a conceptual model of a favorable boundary layer thermodynamic structure for convective downbursts. The ground elevation on the panhandle increases from near 2000 feet at 100W longitude to nearly 5000 feet at 103W longitude (Oklahoma Climatological Survey 1997). Atkins and Wakimoto (1991) discussed the effectiveness of using mesonet observation data in the verification of the occurrence of downbursts. Correlation between GOES MWPI values and observed surface wind gust velocities was computed to assess the significance of a linear relationship between observed downburst wind gust magnitude and MWPI values. Next Generation Radar (NEXRAD) base reflectivity imagery (level II) from National Climatic Data Center (NCDC) was utilized to verify that observed wind gusts were associated with convective storms. NEXRAD images were generated by the NCDC Java NEXRAD Viewer (Available online at http://www.ncdc.noaa.gov/oa/radar/jnx/index.html). Another application of the NEXRAD imagery was to infer microscale physical properties of downburst-producing convective storms. Particular radar reflectivity signatures, such as the bow echo and rear-inflow notch (RIN)(Przybylinski 1995), were effective indicators of the occurrence of downbursts.

## 3. Case Studies

**29 March 2007 Downbursts**

During the evening of 28 March 2007, strong downbursts were observed over the Oklahoma Panhandle. Downburst wind gusts of 45 and 47 knots were recorded by the Goodwell and Beaver, Oklahoma mesonet stations at 0055 and 0350 UTC 29 March 2007, respectively. Late afternoon (2200 UTC 28 March 2007) Geostationary Operational Environmental Satellite (GOES) Microburst Windspeed Potential Index (MWPI) imagery indicated elevated values along and to the east of the dryline that was located over the western Oklahoma Panhandle. Elevated MWPI values indicated favorable boundary layer thermodynamic conditions for convective downbursts most likely resulting from the presence of the dryline. The dryline is defined as a narrow zone of extremely sharp moisture gradient that separates moist air originating over the Gulf of Mexico from dry air originating from the semi-arid high plateau regions of Mexico and the southwestern United States (Schaefer 1986). Stull (1988) has identified that the characteristics of a deep convective mixed layer are caused by a combination of buoyant heat flux, due to strong solar heating of the surface, and wind shear. These conditions are typically found along the dryline. Also, Ziegler and Hane (1993) found the presence of a deeper, well-mixed CBL in proximity to the dryline. In addition, increased vertical circulation, resulting from the sharp temperature and moisture gradients along the dryline zone, is believed to be responsible for enhanced mixing. The dryline established favorable conditions for downbursts by enhancing vertical circulation and hence, the depth of the CBL.

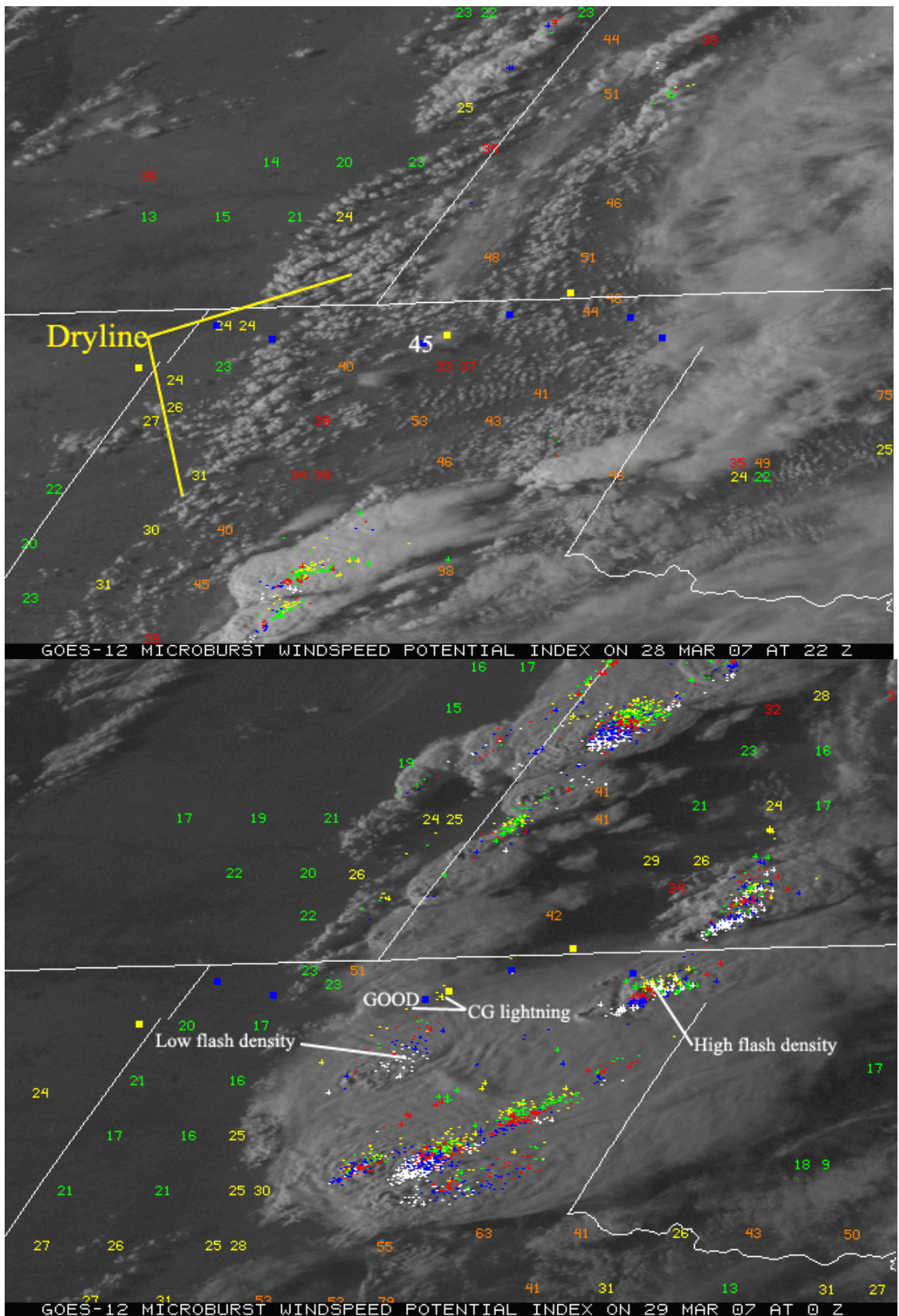

Figure1. GOES MWPI product images at 2200 UTC 28 March 2007 (top) and at 0000 UTC 29 March 2007 (bottom).

Figure 1, GOES MWPI product images at 2200 UTC 28 March 2007 and at 0000 UTC 29 March 2007, respectively, display the development of convective storm activity over the Texas and Oklahoma Panhandles.  The 2200 UTC image indicates elevated values (plotted in red and orange) along and to the east of the dryline.  The 0000 UTC image displays cloud-to-ground (CG) lightning activity near the time of downburst occurrence at Goodwell (45 knots at 0055 UTC).  Noteworthy is a lack of CG lightning in the vicinity of Goodwell at the time of downburst occurrence as well as decreased CG lightning flash density associated with convective storms near the dryline.  Pryor (2006b) described the role of an elevated charge dipole in the appearance of lightning "gaps" associated with convective downbursts.  An elevated dipole most likely effected suppressed CG lightning activity associated with the downburst observed at Goodwell as evidenced by storm radar echo top heights near 45000 feet (not shown), well in excess of the height of the -20C isotherm at 20000 feet as indicated in a Rapid Update Cycle model (RUC) sounding (not shown).  Pryor (2006b) noted a height difference between the -20C isotherm and the storm echo top greater than 25000 feet associated with an elevated dipole.

Inspection of the RUC sounding profile over Goodwell, revealed an elevated mixed layer and inverted V profile between 670 and 750 mb.  Below 750 mb, strong southerly flow and associated moist air advection resulted in a nearly constant dewpoint depression with height.  The elevated mixed layer depth is more typical of the warm-season environment over the High Plains.  Overall, the sounding profile, similar to the type A (Wakimoto 1985), reflected the favorability for downbursts as indicated in GOES MWPI imagery at 0000 UTC.

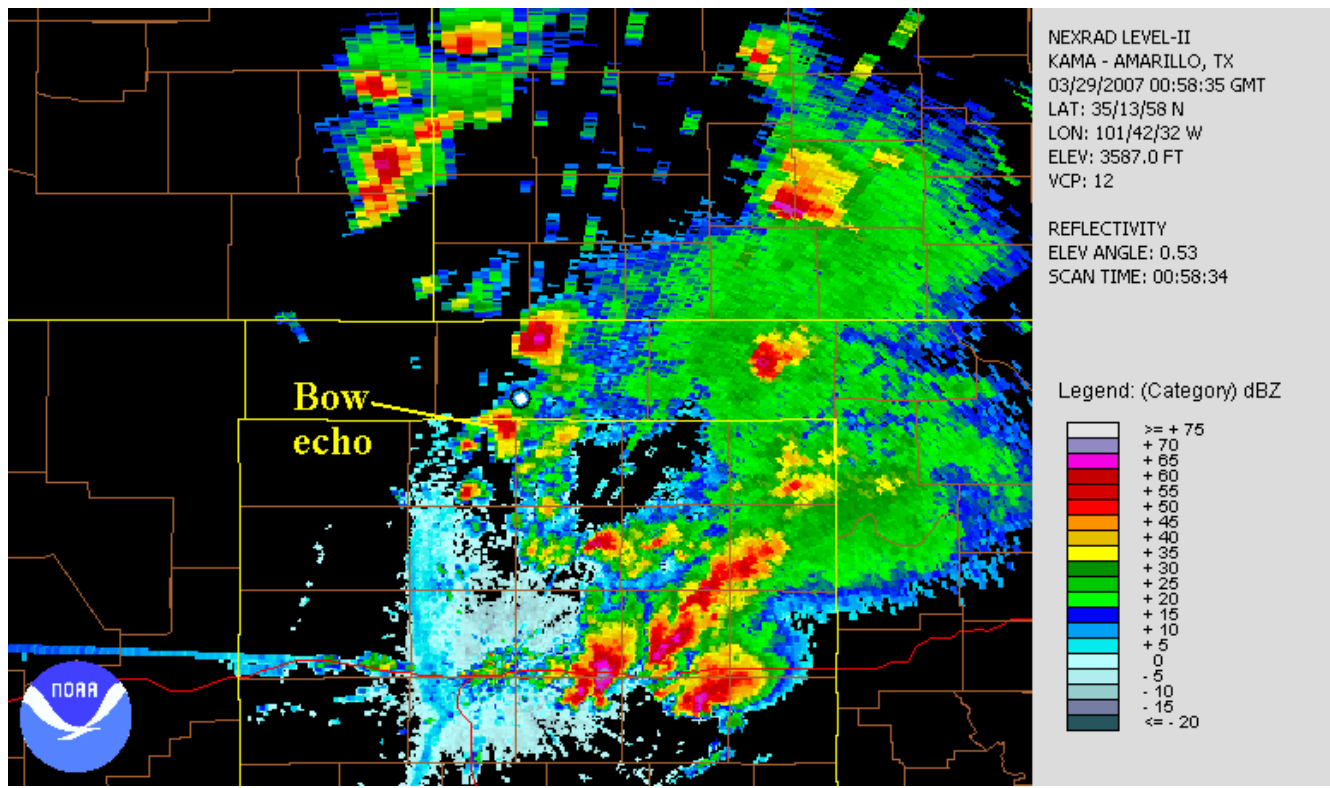

Figure 2.  NEXRAD reflectivity image at 0058 UTC 29 March 2007.  Marker indicates location of Goodwell mesonet station.

Radar reflectivity imagery in Figure 2 was particularly effective in the verification of the occurrence of the downburst observed at Goodwell.  The Amarillo, Texas and Dodge City, Kansas NEXRADs identified bow echoes southwest of the Goodwell mesonet station and southeast of the Beaver mesonet station (not shown) at time of observance of downburst wind gusts.  Przybylinski (1995) discussed the importance of the bow echo signature as an identifier of downburst occurrence in which the strongest downburst winds are expected to be observed near the bow echo apex.

Figure 3, a meteogram from the Goodwell, reflected downburst occurrence by displaying a sharp peak in wind speed at approximately 0055 UTC.  Atkins and Wakimoto (1991) related this peak in wind speed to downburst observation at the surface by mesonet stations.  Interesting to note was an increase in MWPI values downstream of the convective storm activity through the evening in the region of downburst occurrence.  Thus, MWPI product imagery indicated favorable environmental conditions for downburst occurrence associated with the convective storms that tracked through the Oklahoma Panhandle during the evening of 28 March.  In addition, radar reflectivity imagery and surface observation histograms (meteograms) provided essential ground truth in the GOES MWPI validation process.

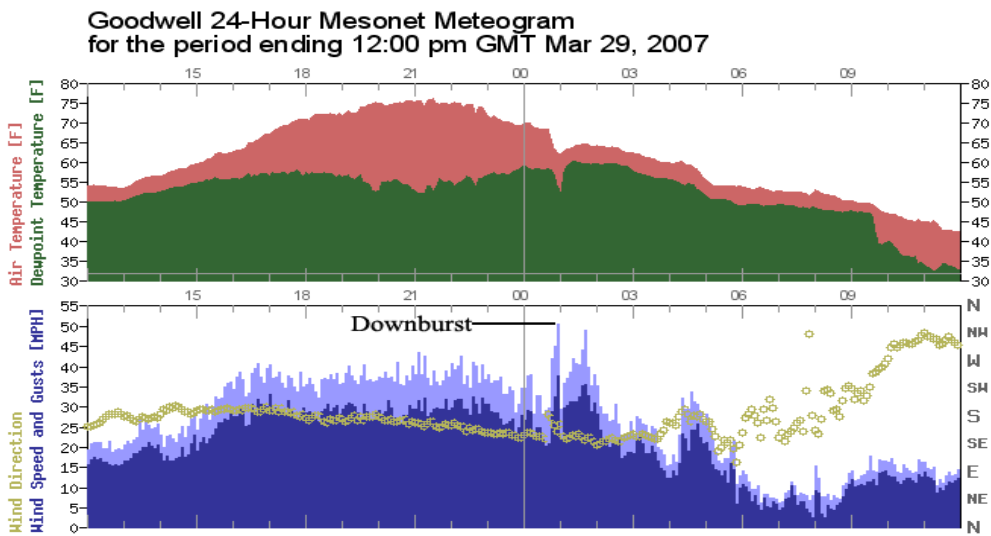

Figure 3.  Oklahoma Mesonet meteogram at Goodwell.

**10 April 2007 Downbursts**

During the afternoon of 10 April 2007, strong downbursts, associated with convective storms initiated by an upper-level disturbance, were observed over the western Oklahoma Panhandle.  The first and strongest downburst wind gust (44 knots) was recorded by the Kenton mesonet station where the highest MWPI value in the panhandle was indicated over an hour earlier at 2000 UTC.  This event demonstrated a strong correlation between MWPI values and observed convective surface wind gusts.  In a similar manner to the previous case, a lack of CG lightning was noted at the time and location of downburst occurrence.

Analyses of surface observations (not shown) from the Oklahoma Mesonet revealed that a strong cold front tracked through the panhandle region between 1400 and 1700 UTC.  By 2000 UTC, the cold front was apparent in visible satellite imagery as a line of enhanced cumulus over the Texas Panhandle.  Strong surface winds and solar heating of the boundary layer after the passage of the cold front resulted in the development and evolution of a deep convective mixed layer through mid-afternoon, establishing favorable conditions for downburst generation.  Favorable environmental pre-conditioning for downbursts was apparent in 2000 UTC MWPI imagery in Figure 4 with the highest risk values located over Cimarron County, Oklahoma. Also relevant in the imagery is the developing cluster of convective storms over southeastern Colorado that would eventually propagate southeastward over the Oklahoma Panhandle. The GOES sounding from Clayton, New Mexico at 2000 UTC in Figure 5 echoed a favorable environment for downbursts by displaying an inverted V profile, similar to the type A profile as described by Wakimoto (1985).

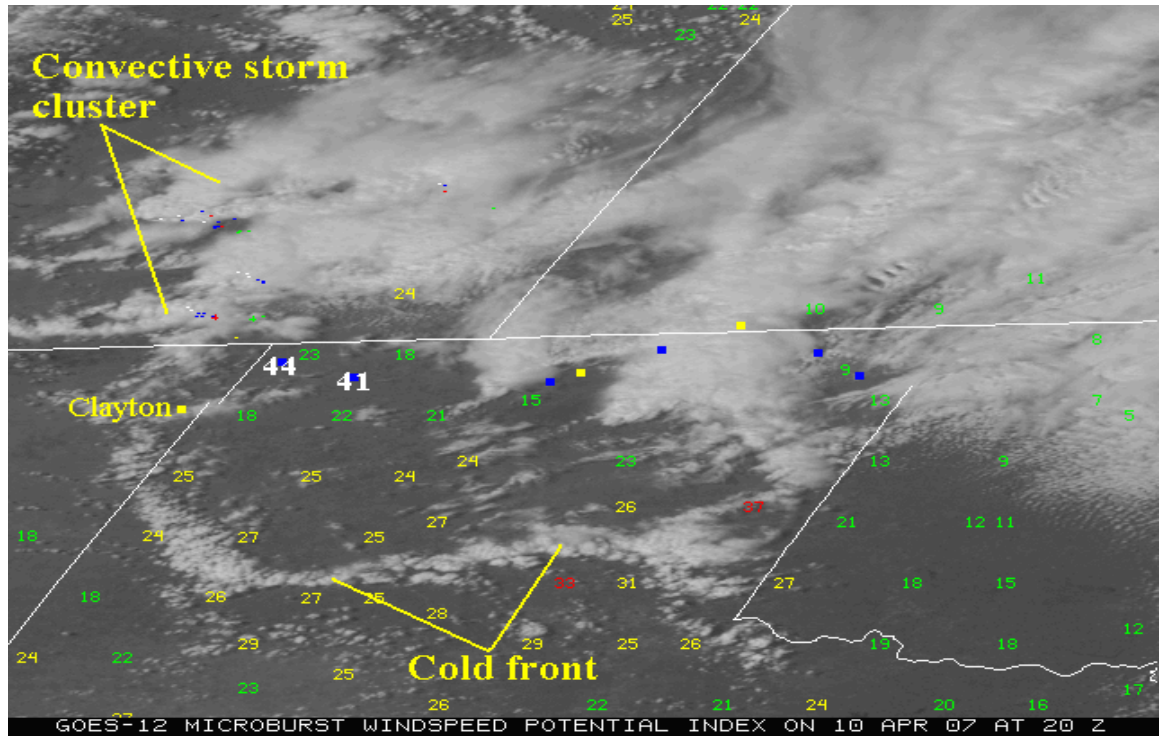

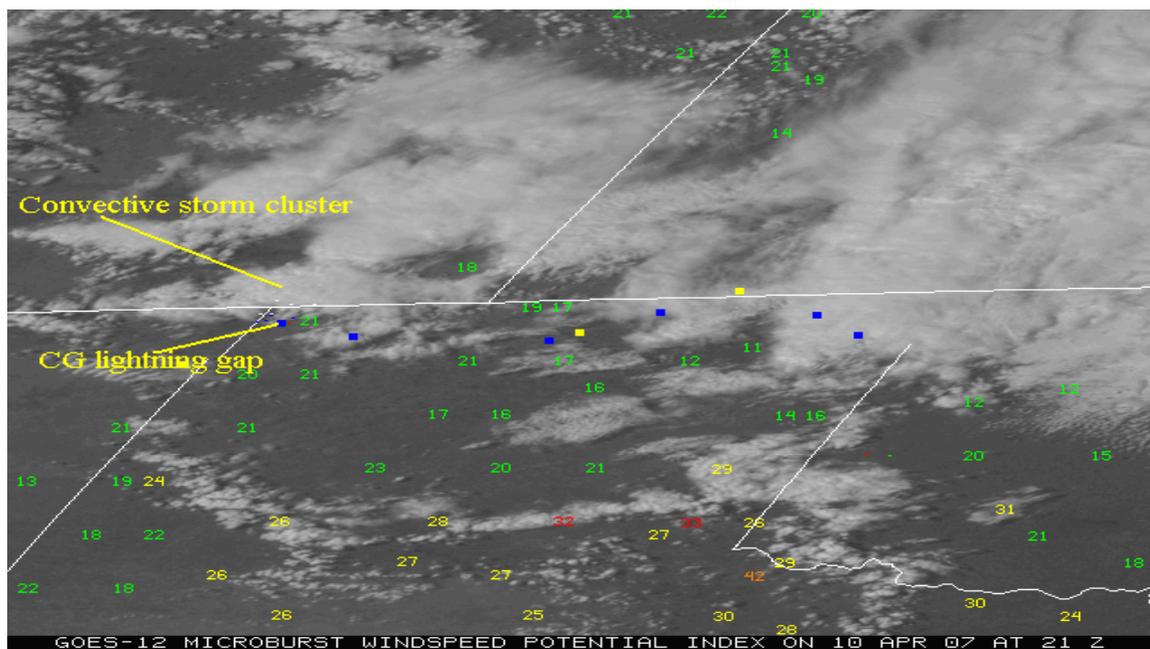

Figure 4. GOES MWPI images at 2000 UTC (top) and 2100 UTC (bottom) 10 April 2007.

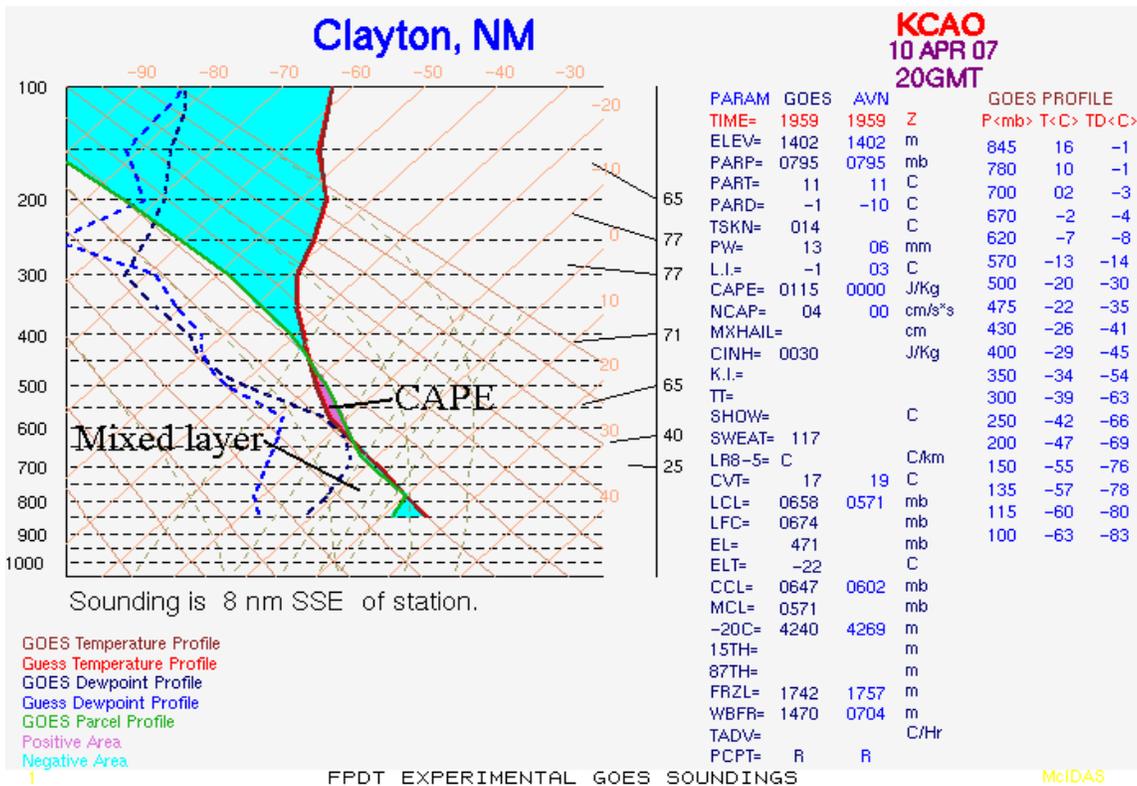
Figure 5.  GOES sounding profile at 2000 UTC 10 April 2007.

Radar reflectivity imagery from Amarillo, Texas NEXRAD in Figure 6 displayed the evolution of convective storm activity as the storm cluster tracked through the Oklahoma Panhandle.  Similar to the previous case, bow echo signatures were apparent at the time and location of downburst occurrence at both the Kenton and Boise City mesonet stations.  Lower reflectivities and storm echo top heights (at or below 25000 feet, not shown) than indicated in the previous case associated with the downburst-producing storms were also evident in radar imagery.

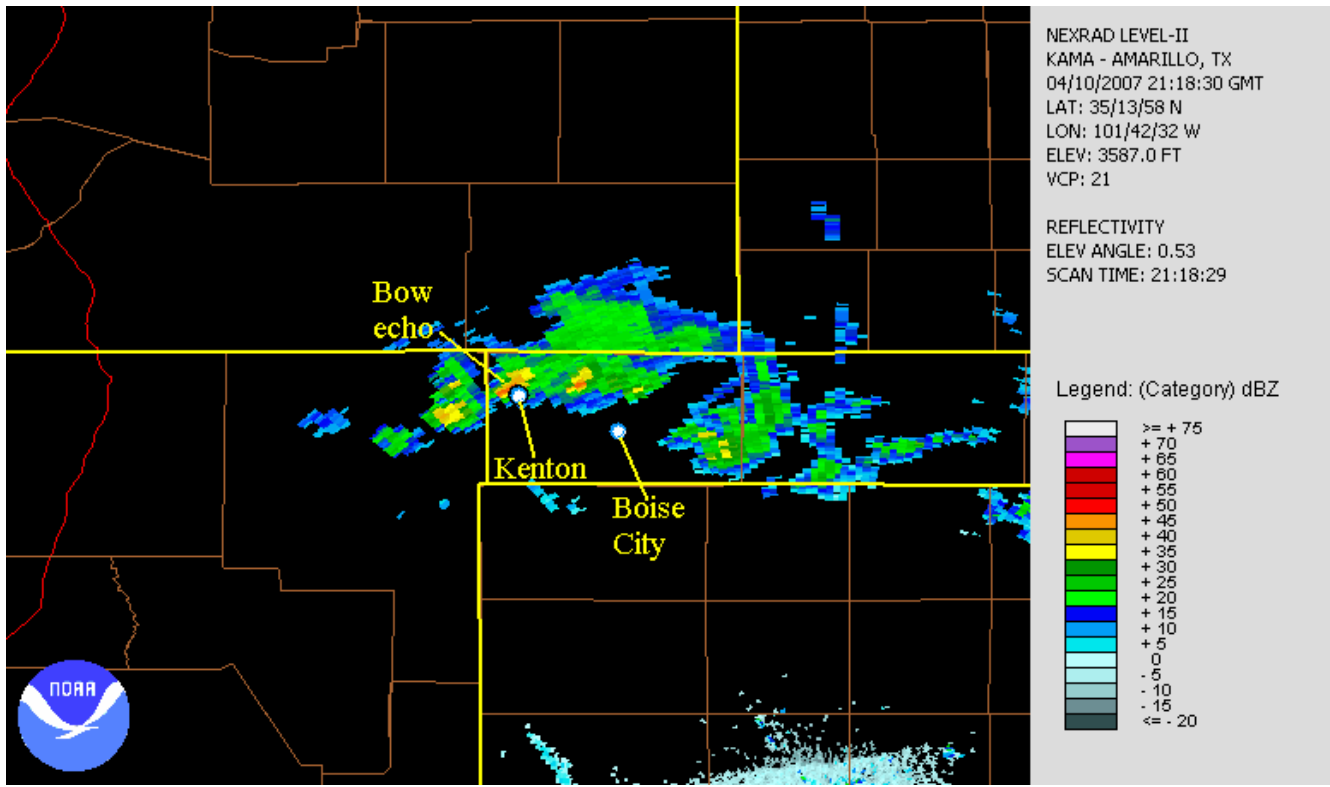

Figure 6. NEXRAD reflectivity image at 2118 UTC 10 April 2007. Markers indicate locations of Kenton and Boise City mesonet stations.

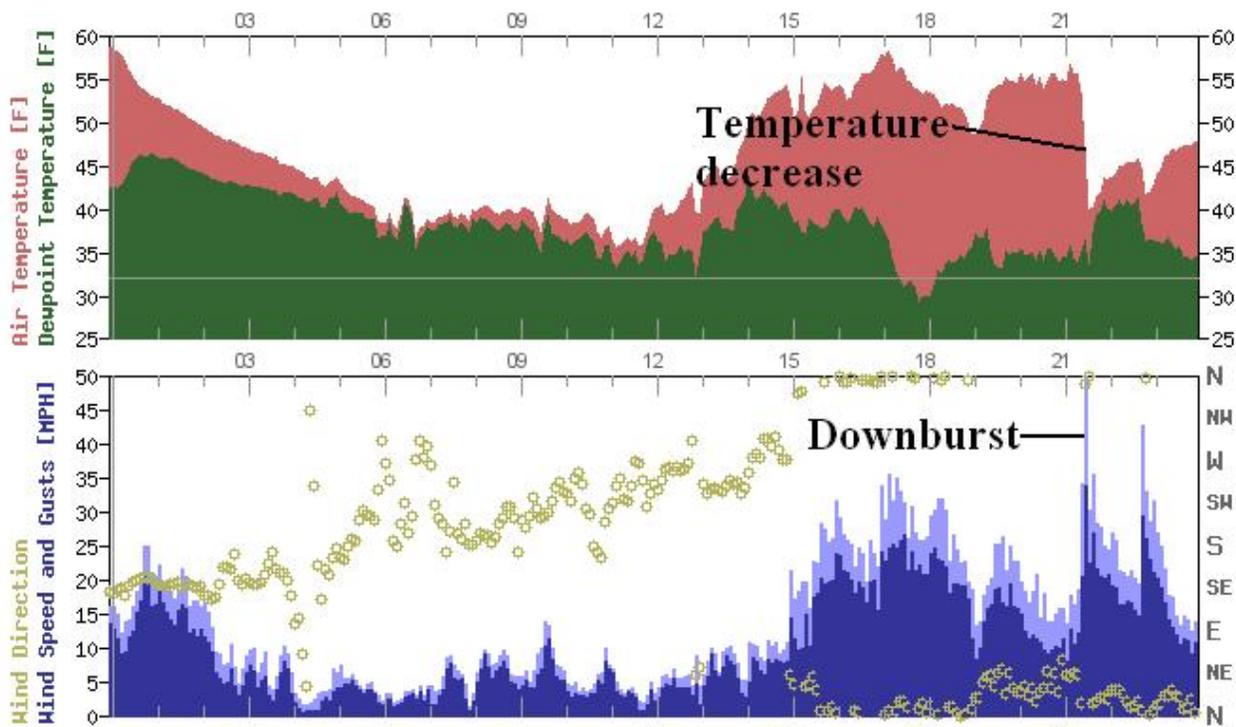

Figure 7. Oklahoma Mesonet meteogram at Kenton.

The meteogram shown in Figure 7 from Kenton indicated downburst occurrence at 2125 UTC. Sharp peaks in wind speed and temperature decreases confirmed downburst occurrence in this study. Downburst gust magnitude decreased from 44 knots at Kenton to 41 knots at Boise City, corresponding closely to the decrease in MWPI values as indicated in the 2000 UTC image. MWPI imagery in Figure 4 also displayed the progress of the convective storm cluster as it entered the panhandle between 2100 and 2200 UTC. Readily apparent in the imagery is a CG lightning "gap" associated with the downburst observed at Kenton and no CG lightning associated with the downburst-producing storm observed at Boise City. In contrast to the previous case, a deficiency of mixed phase precipitation between the levels of the -10C and -20C isotherms and storm echo top heights that did not greatly exceed the level of the -20C isotherm (near 14000 feet) were most likely factors in reduced CG lightning discharge (Saunders 1993) with these downburst-producing storms. MWPI imagery, used in conjunction with radar imagery and surface observations from the Oklahoma Mesonet, demonstrated utility in the short-term prediction of downburst magnitude and downburst verification.

### 4. Conclusions

Validation (see appendix) based on two early spring downburst events over the Oklahoma Panhandle indicated a strong correlation ($r = 0.93$) between MWPI values and observed surface convective wind gusts. As exemplified in the case studies, the GOES MWPI product demonstrated utility in the short-term prediction of downburst magnitude. Future validation effort will focus on upcoming warm season (June-August) downburst events that occur over the High Plains, specifically the Oklahoma Panhandle region. In addition, as outlined in Pryor (2006b), cloud-to-ground (CG) lightning data from the National Lightning Detection Network (NLDN) will be analyzed to investigate and derive a relationship between CG lightning spatial patterns and the location of peak downburst wind gusts.

**Acknowledgements**

The author thanks Mr. Derek Arndt (Oklahoma Climatological Survey) and the Oklahoma Mesonet for the surface weather observation data used in this research effort. The author also thanks Jaime Daniels (NESDIS) for providing GOES sounding retrievals displayed in this paper. Cloud-to-ground lightning data was available for this project through the courtesy of Vaisala, Inc.

**Appendix**

**Correlation:**

| | | | | | |
|---|---|---|---|---|---|
| MWPI to measured wind: | 0.93 | MWPI to HMI: | 0.94 | Mean HMI: | 22.25 |
| HMI to measured wind: | 0.99 | MWPI to ET: | 0.92 | Mean MWPI: | 28.5 |
| | | No. of events: | 4 | Mean Wind Speed: | 44 |

| Date | Time | Measured Wind Speed kt | Location | GOES-12 HMI | GOES-12 MWPI | Max ET |
|---|---|---|---|---|---|---|
| **29-Mar-07** | 0:55 | 45 | Goodwell | 24 | 35 | 40000 |
| | 3:50 | 47 | Beaver | 26 | 38 | 45000 |
| **10-Apr-07** | 21:25 | 44 | Kenton | 22 | 23 | 20000 |
| | 21:45 | 41 | Boise City | 17 | 18 | 25000 |